\documentclass[twocolumn,prl,superscriptaddress,floatfix]{revtex4}
\usepackage[utf8]{inputenc}
\usepackage{amsmath}
\usepackage{amssymb}
\usepackage{graphicx}

\makeatletter
\usepackage{graphics}
\usepackage{epsfig}
\usepackage{color}

\newcommand{\be}{\begin{equation}} \newcommand{\ee}{\end{equation}}
\newcommand{\bea}{\begin{eqnarray}} \newcommand{\eea}{\end{eqnarray}}

\makeatother

\begin{document}

\title{Effect of surface pinning on magnetic nanostuctures}
\author{Aditi Sahoo}
\author{Dipten Bhattacharya}\affiliation{Advanced Mechanical and Material Characterization Division, CSIR-Central Glass and Ceramic Research Institute, 196, Raja S.C Mullick Road, Kolkata, 700032 India}
\author{P. K. Mohanty} \affiliation{CMP Division, Saha Institute of Nuclear Physics, HBNI, 1/AF Bidhan Nagar, Kolkata, 700064 India}
 
\date{\today}

\begin{abstract}
Magnetic nanostructures  are  often considered  as  highly functional materials because they exhibit unusual magnetic properties  under different external conditions. We study the effect of surface pinning on  the core-shell magnetic nanostuctures  of different shapes and sizes considering the spin-interaction  to be Ising-like.  We  explore  the hysteresis  properties and find that  the 
exchange bias, even  under  zero field cooled conditions,  increases with increase of, the  pinning density and the fraction of up-spins among the pinned ones.  We explain these behavior  analytically by introducing a simple model of the surface. The asymmetry in  hysteresis is  found to be  more   prominent in  a inverse core-shell structure, where  spin  interaction  in the core is  antiferromagnetic and that in the shell is ferromagnetic.  These studied   of inverse core-shell structure  are  extended to   different  shapes, sizes, and different spin 
interactions, namely  Ising,  XY-  and Heisenberg models in three dimension. We also briefly discuss the pinning effects  on  
magnetic heterostructures. 
\end{abstract}
\maketitle

\section{Introduction}

In recent few years, nanomaterial has become one of the  interesting field of research \cite{book1, book2} as it shows important novel properties in mesoscopic scale \cite{mesoscopic}.  The nanoparticles  have been  in the center of attention of researchers  for many years as the change of scale  from micro- to nanometer increases  the surface to volume ratio leading to a drastic change in the chemical and physical properties of the system \cite{book3}. Nano-particles can be  synthesized with different spatial structures ,  like  nano-composites, heterostructure and core-shell structure, etc. \cite{book2, Khan, Indra}.

Heterostructures are layered composites prepared from  combination of  materials \cite{2DnanoHetero, Hetero}  having  different physical properties like ferromagnetic-antiferromagnetic, superconductor-ferromagnetic, semiconductor-ferromagnetic, etc.. These combination produces a rich  variety  of novel physical phenomena  strikingly  different from   the individual constituents. The  examples   include   large exchange bias \cite{Bhattacharya, Heblar}, giant magneto-resistance \cite{Song},  varied  interlayer exchange coupling \cite{Tivakornsasithorn}, unusual  spin transport \cite{Garcia}.

Another   interesting  structure is a core-shell structure  which  is found naturally \cite{naturally}, or  can be produced as a  nano-particles  in the lab \cite{in_lab, in_lab2}. Study of core-shell nanostructures   have become popular  recently  as they find  strong potential applications in different fields of research like biomedical \cite{Laurent, BioMed}, catalysts \cite{Zhang}, memory device application \cite{Lo}  etc.,    as their magnetic properties can be varied easily by changing    the size, shape, core or  shell thickness, or other   intrinsic properties.  These materials can be used for  magnetic switching \cite{Sudfeld}, magnetotransport \cite{Kinsella}, micromechanical sensors \cite{Goubault}, DNA separation \cite{Tabuchi}, magnetic memory device\cite{Jia}. bio-imaging and  magnetic resonance imaging  enhancement \cite{Wang}, targeted drug delivery \cite{McCathy}, etc..

The core-shell structure  can be of  different types : (i) magnetic core-non magnetic shell \cite{Xuan}, (ii)magnetic core-ferri or anti-ferromagnetic shell \cite{Vatansever}, (iii) magnetic core-plasmonic shell \cite{Kwizera}, (iv) polymar core-semiconductor shell \cite{Rong}, (v) metallic core-metallic shell \cite{Li}.  The unusual magnetic properties observed  in  all these different structures   owes  to  their mesoscopic scale,  where size and shape-anisotropy play an important  role.  It is known that the change in preferential orientation of magnetic easy-axis is responsible for  changing the value of coercivity \cite{Shirsath}.  The  microscopic  origin of the  exchange bias   however  depends  on many  details, like the   thickness  of core and shell \cite{thickness},  anisotropic spin interaction\cite{Chandra}  and interaction in the core-shell interface \cite{Iglesias}.  In addition,  pinning of spins at the  interface \cite{Ong} or at the the surface \cite{SurfPin}  can affect the magnetic behavior strongly.

In this  article we study how spin pinning   on the surface affects the  hysteresis   properties of the magnetic  core-shell structure. 
First we consider a  circular core-shell structure   where  Ising spins   on lattice  sites belonging to the  core  interact  antiferromagnetically whereas those in the shell   interact ferromagnetically.  The   interaction   in  the interface can be ferro  or antiferromagnetic  in nature.   Spins on the surface, which  are part of the shell,  also interact ferromagnetically  but  some  of the  these spins  are pinned in a sense that their  orientation  does not  change  with time.   Monte Carlo simulations  of the  core-shell structure  reveals that the   exchange bias strongly depends on  the fraction of  $\uparrow$-spins  among  the pinned  ones, and the  pinning density.  A simple   model  of the surface is introduced  to   understand these   properties. 
We also  investigate  the  magnetic properties  of core-shell structure    of  different  shapes:  elliptical,  square, triangular and  some irregular shapes   and  verify  that  surface pining indeed  affect the  coercivity  and exchange bias  strongly.  For continuous spin variables, in XY-   or Heisenberg models,   the  magnetic transitions  can occur  only in three  dimension.  We compare the magnetic properties   of  core-shell nanostructure in three  dimension, with  different spin interactions following  Ising, XY and Heisenberg models. These studies  are also extend  to magnetic   heterostructures, where    spins  in the  bottom layers interact  antiferromagnetically  whereas  interaction  in the  top  layers  is   ferromagnetic.  Interestingly, we find that  the coercive   field in  the  reverse direction  of the hysteresis loop   can become positive  when   the   number of ferromagnetic layers  are decreased below a threshold value.

 \section{The Model}
 We intend to study, primarily,  the  magnetic  properties   of  core-shell nanostructures. We assume that 
the external temperature is smaller than  the critical  threshold $T_c$  above which  the magnetic order disappears.  Simple models  which undergo magnetic transition  at  some 
 finite temperature $T_c\ne 0,$  are   Ising model in two and three  spatial  dimensions ($d=2,3$), XY model in three dimension ($d=3$) and  
 Heisenberg model in three dimension ($d=3$).  This  is because the non-existence theorem \cite{nonExist}  rules out possibility of  finite temperature phase transition in  one dimension ($1d$)  and  the  Mermin-Wagner \cite{MW} theorem   does not allow continuous symmetry, one  like  the  planer  spins in XY  model  or  three dimensional spins with unit magnitude  in   Heisenberg model, to be broken in $d\le 2.$ 
 
 Let us consider a disk-like nanoparticles  in  $d=2$  with constituent spins having Ising  symmetry.  This   is the simplest model that exhibits  intrigue and interesting    magnetic properties in mesoscopic scale  when the surface spins are  pinned.   A generalization  to spherical  particles   in three dimension  with  spins following   Ising, XY or Heisenberg symmetry  is straight forward; we discuss these  cases  briefly in section \ref{sec:3D}.  
 
  \begin{figure}[h!]
 \begin{centering}
 	\includegraphics[scale=0.25]{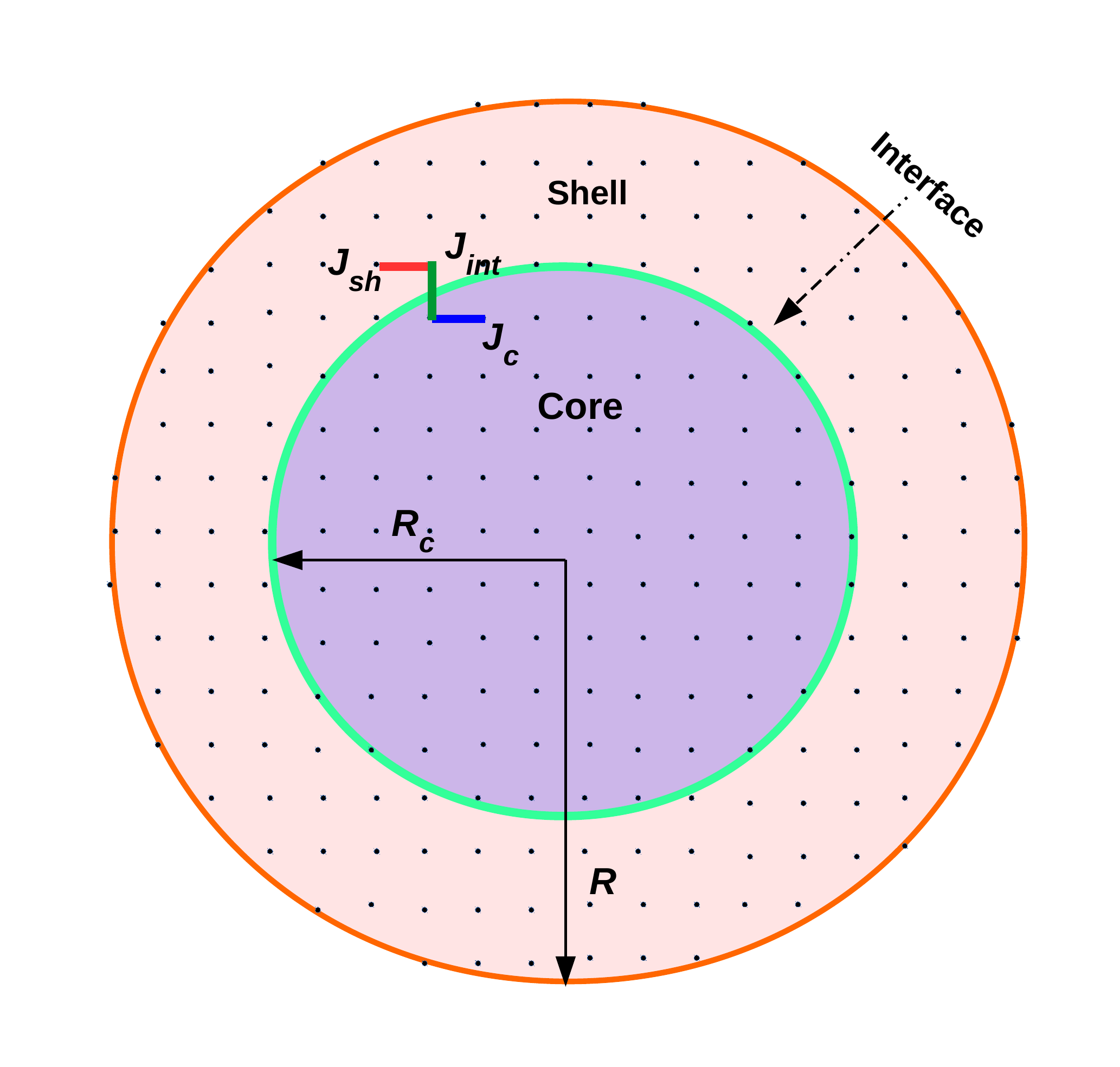}
 	\caption{(Color online) Schematic representation of the core-shell structure:  The core-shell structure   in two dimension  is  represented here by a  square lattice  bounded by a circle  of radius $R$, within which there  is a  circular-core having radius $R_c<R;$  the annular region of width  $R-R_c$ is the shell.  Ising spins $s_i = \pm 1,$ on the  lattice sites belonging to the core (shell)   interact 
 	with strengths  $J_c$ ($J_{sh})$  and the  interaction strength across  the interface is $J_{int}.$     }
 	\label{fig:cartoon}
 	\par\end{centering}
\end{figure}

 We consider a square lattice  with   circular boundaries of radius $R$, 
 which  has a  circular core ${\cal C}$ of radius $R_c<R;$  the shell region  ${\cal S}$ falls  between these two circles (see  Fig. \ref{fig:cartoon}).  Each lattice  site  $i$  of this  core-shell structure  is  associated with  a   Ising spin  $s_i=\pm1$ (representing $\uparrow$, $\downarrow$)  which interact  following the  Hamiltonian, 
\bea
 {\cal H} &=&   -J_c \sum_{i\in {\cal C}, j\in {\cal C}} s_i s_j  -J_{sh} \sum_{i\in {\cal S}, j \in {\cal S}} s_i s_j \cr && -  J_{int}\sum_{i \in {\cal S}  j \in {\cal C}} s_i s_j   
 -H \sum_{i\in {\cal C}, i\in {\cal S} } s_i,
 \label{eq:H}
 \eea
 where $j$ is the nearest neighbor of  site $i,$  $J_c$ ($J_{sh}$)  are  the exchange interaction  among the spins within the core (shell), $J_{int}$   represents  core-shell interface   interaction and  $H$ is the external magnetic field. Obviously, the  spins  interact ferromagnetically (antiferromagnetically) when the corresponding value of  interaction  strength is positive (negative);  for example    when   $J_{c}<0,$  and sites  $i$ and $j$ are within the core ($i \in {\cal C}$ and $j\in {\cal C}$) the spins $s_i$ and $s_j$  interact antiferromagnetically. The magnetic field $H$   is experienced by all the  spins, both  in core  ${\cal C}$  and   shell ${\cal S}$.  In addition, to mimic surface pinning  which has been  observed 
 in  magnetic nanoparticles   in contact of  organic  liquids  \cite{Berkomitz}, in  ferromagnetic thin films \cite{Sparks} and  core-shell structures \cite{SurfPin} etc.,  we assume  that $\eta$-fraction of boundary spins are pinned; a parameter $0\le r\le 1$  controls the fractions  of $\uparrow$ spins among  pinned  ones.    
 
We  proceed  by assuming  the interaction of spins   within the core is  antiferromagnetic ($J_{c}<0$) whereas the  interaction of spins   in the shell   is  ferromagnetic ($J_{sh}>0$) and  the interaction   at the interface   can be ferro-  or antiferromagnetic; this  is not the usual scenario  but observed   in   several  experimental systems \cite{Vasilakakz}. We briefly discuss other  possible structures in 
section \ref{sec:AFshell}. In the following we  primarily  study the  hysteresis effects,  particularly    the dependence  of    coercive field and exchange bias  change on different model parameters and the  shape  and size of the core-shell  structure.

\section {Core-shell structures  in two dimension}
\subsection{Anti-ferromagnetic core and  Ferromagnetic shell}
We start  with   nanoparticles  having a circular core-shell structure  with radius $R_c$ and $R$  respectively.  The interaction  among 
spins   is described by  the Hamiltonian  Eq. (\ref{eq:H}), with  three exchange coupling constants $J_c, J_{sh}, J_{int}$ and  the  external field  $H.$ We  proceed to do the Monte Carlo  (MC) simulations of the model at temperature  $\beta^{-1} =1$ and   pinning density  $\eta.$  Thus,     the total number of   pinned  spins   is $N=\eta L$ where  $L$ is number of lattice  sites on the surface - these  $N$ spins are chosen out of $L$  randomly and independently.   Further,   we choose $rN$  spins  out of  $N$ randomly and  orient them $\uparrow$; the rest  
of them points $\downarrow.$ 

\begin{figure}[h!]
	\begin{centering}
		\includegraphics[scale=0.25]{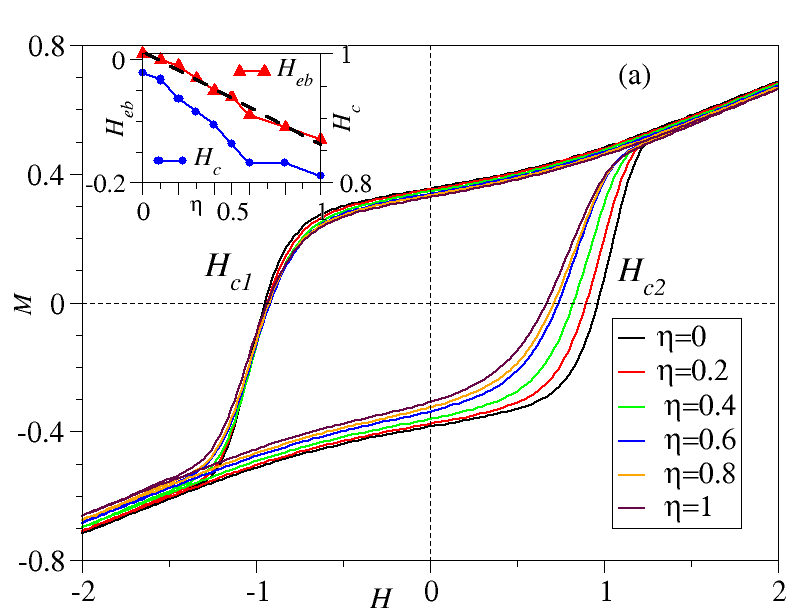}
		\includegraphics[scale=0.25]{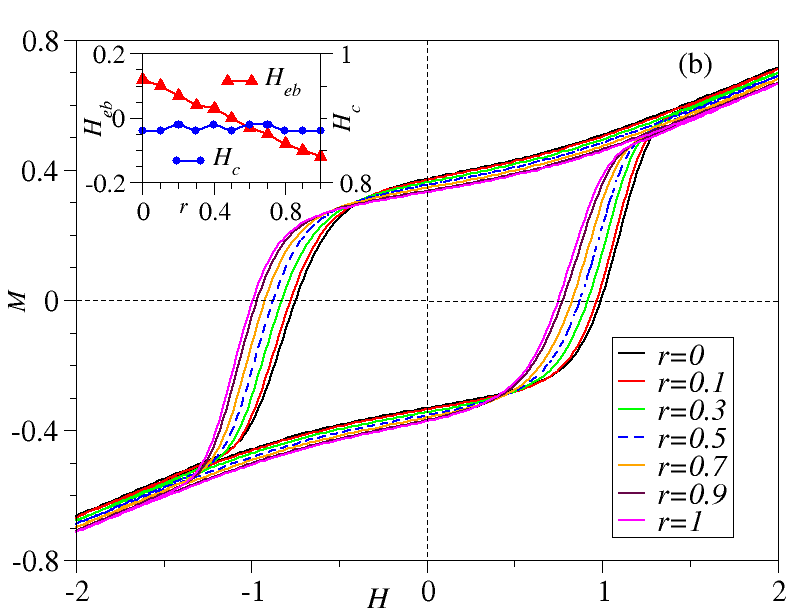}
		\includegraphics[scale=0.25]{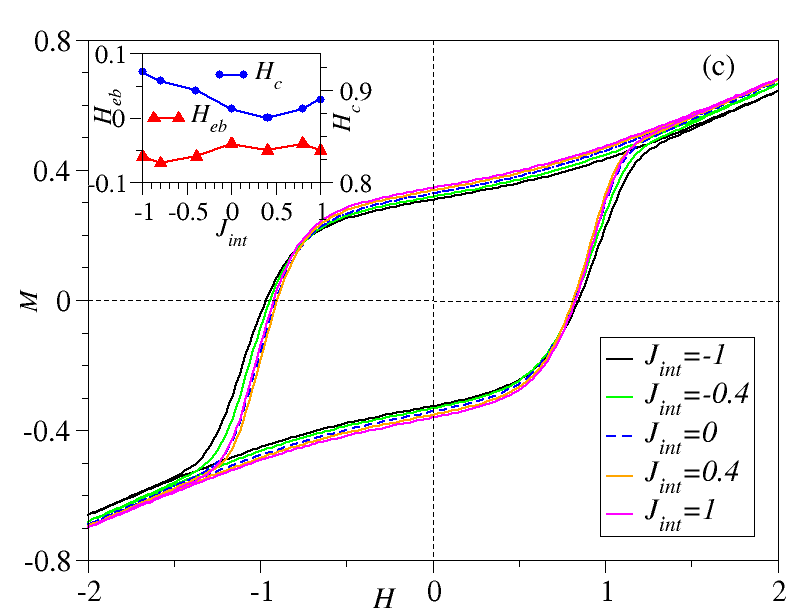}
		\par\end{centering}
		\caption{ (Color online) The  main figures   show hysteresis loops for two-dimensional   circular core-shell structure when  one of the model parameters $X$ is changed  while other parameters are taken from Eq. (\ref{eq:default});  insets show  the corresponding  change in coercivity $H_{c}$ and exchange bias $H_{eb}.$
		(a) $X= \eta:$  $H_{c}$ decreases,  but   $|H_{eb}|$ increases with  increase of $\eta.$ The dashed line  is  
		the best linear fit $H_{eb}= -0.15 \eta +0.013.$ 
		(b) $X= r:$  $H_{c}$  is almost a constant,  but   $|H_{eb}|$ increases linearly  with  increase of $\eta.$
		(c) $X= J_{int}:$  No appreciable  change in  $H_{c}$ and  $|H_{eb}|.$}
		
		\label{fig:eta_and_r}
\end{figure}

We start the  simulation  from a random configuration where all  spins in the  core and the shell, except those which are pinned, 
are  chosen to be $\uparrow$  ($+$) or $\downarrow$  ($-$)  with equal probability, that mimics   an  Ising  configuration  at  infinite temperature and  zero magnetic field. 
In this   zero-field  condition,  we set the   temperature  of the system $\beta^{-1}=1$ and  keep it  fixed throughout the simulation \cite{note1}. To obtain hysteresis 
we  first  raise  the    magnetic field  slowly   from $H=0$  to $H_{max}=2$  with a rate $0.02$ units  per Monte Carlo sweep (MCS)  and  finally   the hysteresis loop calculations are undertaken for a cycle, by varying  the  field from $H_{max}$ to 
$ -H_{max}$  and then to   $ H_{max}$ again with same rate.  Magnetization of the system is measured after each MCS  and  
it is averaged over  $100$ samples.  We primarily focus on  the  dependence of  coercive field $(H_c)$ and exchange bias field $(H_{eb})$ on  pinning density $\eta$,  $\uparrow$-spin fraction $r$,  interface interaction strength $J_{int}$ and core-shell size     $R_c,R.$  

The coercivity  $H_c$ and exchange bias  $ H_{eb}$ in a hysteresis loop are  defined as 
\be
H_{c} =\frac12 (H_{c2}-H_{c1})  ~~ {\rm and } ~~ H_{eb} = \frac12 (H_{c2}+H_{c1}), \label{eq:defn_HcHeb}
\ee 
where $H_{c2}$ and $H_{c1}$ are the fields corresponding to zero magnetic moment in the forward and reverse branches of the loop 
(as shown in Fig. \ref{fig:eta_and_r}(a)). 
Usually   the   coercive   field $H_{c1}$  is  negative and $H_{c2}$ is positive.
The exchange bias, however, can change sign; it is positive (negative)  when   $|H_{c1}| < |H_{c2}|$  ( $|H_{c1}| > |H_{c2}|$). In  ordinary of   asymmetry   in the hysteresis.  In the following we  see that, surface pinning can  produce   non-zero   exchange bias in  core-shell magnetic nanoparticles.

Now, we study  how  hysteresis   properties change when one of the  model parameters changes, 
while the others are taken from the following   default  values, unless  otherwise specified.
\bea
R&=& 32, R_c= 26,\eta=0.4,r=0.7,\cr
J_c &=& -0.5, J_{sh} = 1,  J_{int} = 1. \label{eq:default}
\eea
 
{\it  Dependence of $H_{eb}, H_c$  on $\eta:$ }  First we study the  hysteresis properties    of a nanoparticle   by changing   the pinning density $\eta$. Other  parameters  are kept fixed at the default values given in Eq. (\ref{eq:default}).  
In Fig. \ref{fig:eta_and_r}(a) we present hysteresis loops for different $\eta.$   These loops  show   negative exchange bias which 
increases  as  the pinning density $\eta$ is  increased. The exchange bias is maximum when all the spins are pinned at the surface. 
The reason of the asymmetry in the loop is that 70$\%$  of the pinned  spins  are  $\uparrow$  and thus  one requires some  additional 
magnetic field  to completely reverse the magnetization. 
The coercivity, however,  decreases    with increase of $\eta.$ This is because,  with  increase of   pinning density  more spins of ferromagnetic shell are pinned and  less number  spins  of the  shell    take part in the ferromagnetic dynamics - which  effectively decrease the  shell width.

 {\it  Dependence of $H_{eb}, H_c$  on $r:$ } 
 Different external conditions  pins  the spins  on the surface differently.  If the cause of pinning is  organic solvent, both the  density 
 of pinned  spins  and the  $\uparrow$-spin fraction  may vary in different solvent conditions.  Here we intend to  change   $\uparrow$-spin fraction $r$  and  investigate the hysteresis properties. 
For $r> \frac {1}{2},$  more     $\uparrow$-spins are  pinned compared to the  $\downarrow$ and one expects that   an effective   positive  intrinsic  field  is generated in the system.  Thus,  one needs  some  additional   negative  external magnetic  field to nullify  this effect, resulting in   a negative exchange bias. Similarly,   a positive exchange bias  is expected for $r< \frac {1}{2}.$
In Fig.  \ref{fig:eta_and_r}(b)  we have plotted   the hysteresis curves  for  different  $r$, keeping  $\eta=0.4$ and   other parameters  same as  that  in Eq. (\ref{eq:default}).  The inset  here shows  dependence of $H_{eb}$ and $H_c$  on $r.$   As expected,  $H_{eb}=0$ for $r= \frac12,$  it is   negative (positive)   for $r>\frac12$ ( $r<\frac12$), and   $|H_{eb}|$   increases  as one  moves away from $r= \frac12.$     The coercivity, which primarily depends on  the  pinning density $\eta,$ is almost independent of $r.$

 {\it  Dependence of $H_{eb}, H_c$  on $J_{int}:$ } Now  we   aim at changing $J_{int},$ the interface interaction strength. 
The hysteresis loops for  a particles   with size $R =32$ and  $Rc =26$  are plotted  in Fig. \ref{fig:eta_and_r}(c) 
With  change of $J_{int}$ 
we do not find  any significant change  in $H_c$  and  $H_{eb}$;  two extreme values $J_{int} = 1, -1$  gives rise to a slightly increased  coercive field, but the exchange bias  changes only a little.  Thus,  it appears that  in a  core-shell magnetic system  the 
exchange bias  can be controlled effectively  by the $\uparrow$-spin pinning fraction $r$ and the  pinning density $\eta$, not by the 
interface interaction  $J_{int}.$

One should note that  some earlier studies   have  reported  a significant  change in  exchange bias  with change in   interface  interaction strength $J_{int}$ \cite{Iglesias}.   These studies, primarily  focus   on    core-shell   structure with ferromagnetic core and antiferromagnetic shell \cite{Nogues, Wu, Eftaxias}, 
modeled  by  the  usual Heisenberg model in three dimension  along with,  additional  magnetic  anisotropy \cite{Wu, Eftaxias}     and  sometimes in presence of a crystal field \cite{Zaim}.  In addition  hysteresis is  studied in  both field cooled and zero-field  cooled conditions; the exchange bias and its change   with  respect to   interface interactions  are  found to be  significant only in   field cooled conditions \cite{Iglesias}. 

In the present study  we  have an inverse core-shell structure, with  an  antiferromagnetic core  and  ferromagnetict shell  modeled by 
Ising spins  in  two dimension and  there  is no  magnetic anisotropy or  any crystal field;  in this simple  case, even in zero-field cooled conditions  we find  a large   exchange bias $H_{eb}$ when surface spins are pinned.  $H_{eb}$ will, of course,  increase  further  in field cooled conditions.   To emphasize that 
surface-pinning indeed causes   large  exchange bias, we  extend the  study  to three dimensional inverse 
core-shell structure  considering  Ising, XY- and Heisenberg models (see section \ref{sec:3D}); in all these cases, under zero field cooled conditions,  the affect of  $J_{int}$ on   $H_{eb}$ are found to  be negligible.

\begin{figure}[h!]
	\begin{centering}			
		\includegraphics[scale=0.3]{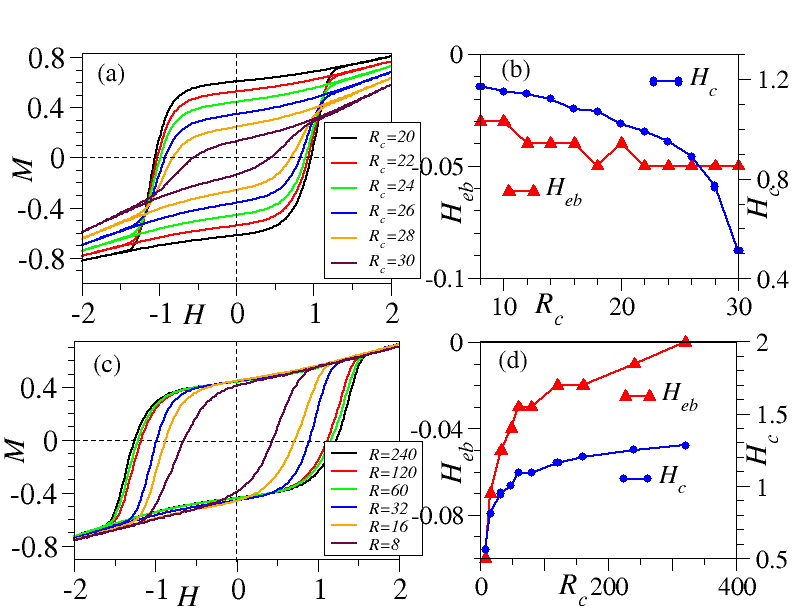}
		\par\end{centering}
	\caption{(Color online) (a) Hysteresis loops  for different  shell thickness $R-R_c,$  obtained  by varying $R_c,$  keeping $R=32$ fixed. 	(b) $|H_{eb}|$ increases  with increase $R_c$ and reach a constant  for large $R_c$. $H_{c}$also  decreases  with  increase of $R_c,$ as the   number of  ferromagnetic  	layers also decrease.  (c) The thermodynamic limit of the system can be  achieved  by changing the size  of  the core and shell  proportionately. Here we plot hysteresis  loops  for different $R,$ and fixed  $\frac {R_c}R = \frac34.$  (d) $H_c, H_{eb}$ as a function of  $R.$   $H_c$ increases with  $R$   as the number of ferromagnetic layers  increase. However  $H_{eb} \to 0$ because 
	the surface (where pinning occurs) to volume ratio   approaches to zero in the  thermodynamic limit. Unspecified parameters here  are taken  from Eq. (\ref{eq:default}). }\label{fig:R_Rc}
\end{figure}

 {\it  Dependence of $H_{eb}, H_c$  on $R$ and $R_c:$ } 
Figure \ref{fig:R_Rc}(a) shows the hysteresis loops of inverted core-shell structure for different $R_c,$  keeping  $R=32$ fixed; thus 
the shell  thickness   increases  with  decrease  of $R_c$.   The other  parameters  are  chosen from Eq. (\ref{eq:default}).
  In  Fig.  \ref{fig:R_Rc}(a) we  plot the     variation of  coercive field   and exchange bias 
with $R_c.$  Here,    $|H_{eb}|$  increases   with increase of $R_c$ and  reaches  a constant value   asymptotically. 
The saturation magnetization  and coercive field, however,  decreases for larger  $R_c.$   This is because,  the coercivity  primarily  gets   contribution from the ferromagnetic shell  (antiferromagnetic core  produces  zero net magnetic moment) whose thickness    decreases   with  increased $R_c.$

 It is important to  ask  whether   the  observed behavior   is scalable, i.e.,  whether the  asymmetric  hysteresis survives  in the thermodynamic  limit where  $R_c$ and $R$  increases  keeping their ratio  fixed.  To study this  we increase  particle size  $R$ 
 while  increasing  the core size $R_c$  proportionately,  $R_c=\frac 34 R.$  Other parameters are  taken as  those in  Fig. \ref{fig:R_Rc}(a).   We  find, in   Fig.  \ref{fig:R_Rc}(c)   that the size of  hysteresis  loop   increase  with $R,$ as the number of ferromagnetic layers  are increased. This is  reflected  in   increased value of   coercivity in Fig.  \ref{fig:R_Rc}(d).  However, the magnitude of the exchange bias  $|H_{eb}|$   decreases   with $R$ indicating the   asymmetry     of the loop   decreases  with $R$ and   one gets a usual 
 symmetric hysteresis in thermodynamic limit.  Thus, the exchange  bias due to surface  pinning  are   only the mesoscopic effects  which
 goes  away  in larger systems when surface to volume ratio  becomes  very small. In fact,  similar  size dependence of coercivity   and $H_{eb}$ has been  observed in system with  ferromagnetic shell and antiferromagnetic core \cite{thickness,Square}.

\subsection{ Ferromagnetic core or Antiferromagnetic Shell \label{sec:AFshell} }
  The core-shell structure that we   studied  so far has  ferromagnetic interaction in the shell and  antiferromagnetic interaction in the core. We have investigated  the other possibilities too. The affects  of surface pinning turned out to be  not  that prominent when the core (shell)  is ferromagnetic (antiferromagnetic)  irrespective of the  interaction in the shell (core).  We decide not to present  these  studies in details as these results   neither add any significant information  nor alter the conclusions of this article.  
  
  When  spins interact ferromagnetically  in the shell, the  pinning of surface spins (which  belong to the  shell)  can produce   
  an effective additional magnetic field which in turn generate an asymmetry in the hysteresis loop.  On the other hand,  when spin interactions in the shell  is antiferromagnetic,  the pinning is less effective as  other  spins in the shell can  orient  in a direction  opposite  to the pinned spins  and  make  an antiferromagnetically ordered  structure throughout; thus, the effective  intrinsic filed produced     in the system is  negligible. In this case some sites 
  might encounter  frustrations and   may give rise to  certain local residual  magnetic moment, but the number  of  the   frustrated spins are statistically very small. On the  other hand,  when  the core  is ferromagnetic , one generally gets a  large hysteresis loops even in absence of  surface pinning  because the volume of the  core  is usually  much larger compared to that of the shell. Thus, the relative change  of exchange bias and coercivity  produced by surface pinning  is  quite small. 
  
   In  summary, surface  pinning  surely affects  the  hysteresis properties   in core-shell nanostructures but the effect is more prominent when   interaction  in the core  is antiferromagnetic and  that in the shell  is ferromagnetic.

\subsection{Different surface morphology}
\begin{figure}[h!]
	\begin{centering}
		\includegraphics[scale=0.28]{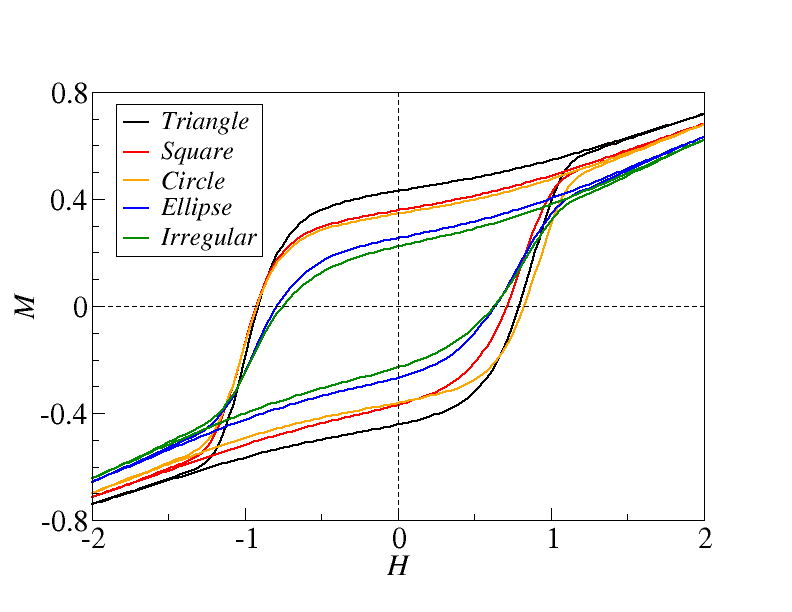}
		\caption{ (Color online) Shape dependence  of   magnetic hysteresis of  core-shell nanoparticles for (i) equilateral triangle (base $a = 86$), (ii) square ($a = 56$),  (iii) circle  ($R=32$), (iv) ellipse (major and minor axis   $a = 46$,  $b =23)$ and (v) an irregular shape (core radius $R_c = 26$). Each one  has  approximately  same  particle size   and the  
		shell thickness. Corresponding  $H_c=0.86, 0.83, 0.89, 0.72,0.7$  and  $H_{eb} = -0.06, -0.11, -0.05,-0.08,-0.06$.  Unspecified parameters are taken as the default values given in Eq. (\ref{eq:default}).} 
		\label{fig:shape}
		\par\end{centering}
\end{figure}

\begin{figure}[h!]
	\begin{centering}
		\includegraphics[scale=0.28]{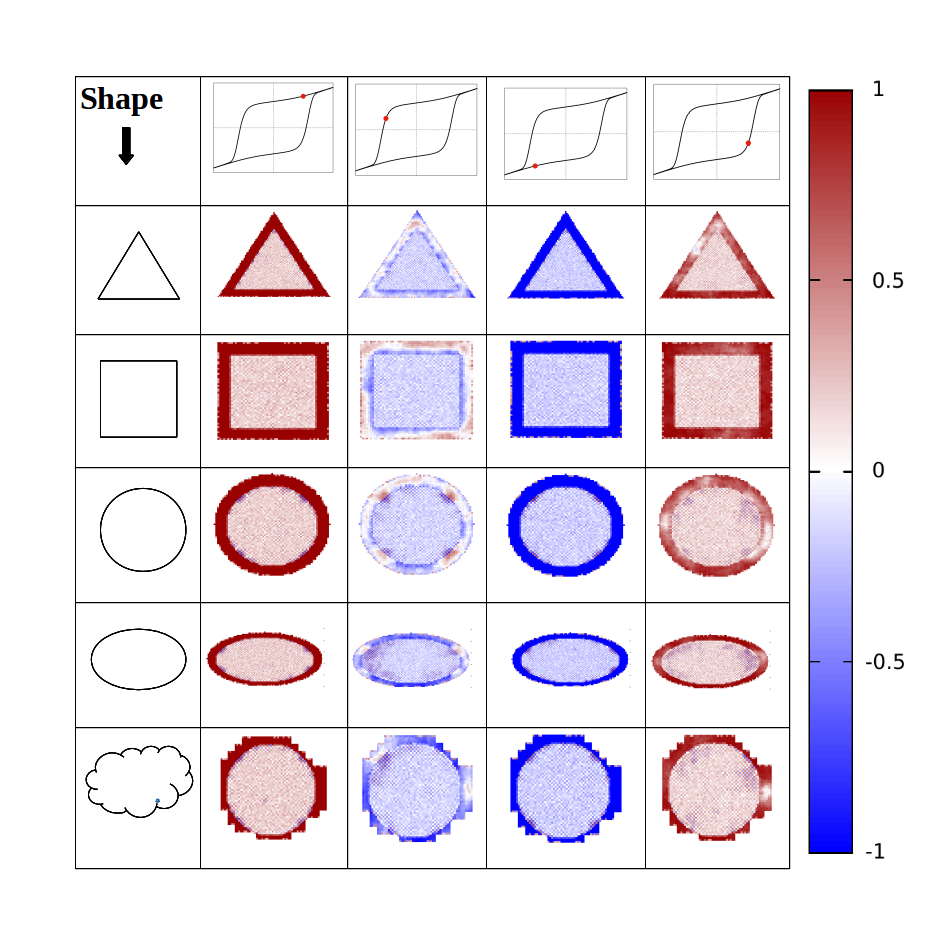}
		\caption{(Color online)  The spin configuration of inverse core-shell structure of different shapes , measured  at four different positions in hysteresis loop, as  indicated  by $\bullet$. Particle size and other parameters are same as in Fig. \ref{fig:shape}.} 
		\label{fig:table}
		\par\end{centering}
\end{figure}

Like other  nanomaterials \cite{diffShape}, the  surface morphology or shape of a core-shell nanocomposites may be changed. Morphology of the nanoparticle can be spherical \cite{Sphere}, squre \cite{Square}, elliptical \cite{Ellipse}, triangular \cite{Triangle} or   it may 
be  irregular \cite{Irregular}.  Their   magnetic properties      depend  crucially  on  the    surface anisotropy \cite{Song2}. Other properties  like catalytic activity, electrical and optical properties are also highly shape dependent \cite{Zwara}. Combination of core-shell materials in different dimensions and shapes are  designed  regularly  for their potential 
application in technology,  like magneto-plasmonic application \cite{Kwizera},  fluorescence application  \cite{Hu}.
In this section  we have studied hysteresis properties   of  two dimensional  core-shell   nanostructures   having  different shapes.

 To emphasize  how   {\it change in surface morphology}   affects  the  magnetic  properties,  we  did   Monte Carlo simulation  of  core-shell  nanoparticles   of  different shapes, but  similar  area and shell-thickness. Coercivity  and exchange  bias  obtained for different shapes  are compared  with    that of the circular   core-shell structure  with $R=32$ and  $R_c=26.$

 We consider  four different shapes, (i)  a triangular core-shell structure with base $a=86$,  (ii) a square core-shell structure with side $a=56$, (iii) an  elliptical   core-shell structure with major axis  $a=46$ and minor axis  $b=23$ and (iv) a  core-shell structure   with irregular surface  but circular core or radius $R_c=26.$  In  all cases except (iv), the shell thickness   is taken   to be $6$ lattice units and for (iv)  the average thickness   is $\simeq 6.$  The   interaction parameters $J_c = -0.5$, $J_{sh} = 1$, $J_{int} = 1$ and the pinning   parameters  $\eta =0.4,$ $r = 0.7$  are  kept same. Hysteresis  loops  of all these different  shapes, along with  that of the  circle,  are plotted in  Fig.  \ref{fig:shape}.  Coercive field of circular shape is found  to be  maximum; then they are  decreasing  in order: circle, triangle, square,  ellipse, and the  irregular shape.   The corresponding   exchange  biases  are $H_{eb}= -0.05, -0.06, -0.11, -0.08$ and  $-0.06$ respectively.

{\it Local magnetic structure :}
The   local magnetic structure  changes during the  hysteresis cycle.  It is interesting to  ask,  how does  the  pinned spins  on  surface of a core-shell nanoparticle   with different morphology,   affects the local magnetic structure. To produce the hysteresis  loop  we  take $H_{max}=2$  and  now look at the   spin configuration  at  four different positions in the hysteresis cycle,  i.e., both in the  forward and backward directions,  at   $H= \pm 1.$  The configurations are   averaged  over $100$ statistical samples to get the  local magnetization  profile  $\{m_i\}  = \{ \langle s_i \rangle\}.$   A density plot  of the magnetization profile is shown  in Fig.  \ref{fig:table}.

\begin{figure}[h!]
\begin{centering}
	\includegraphics[scale=0.28]{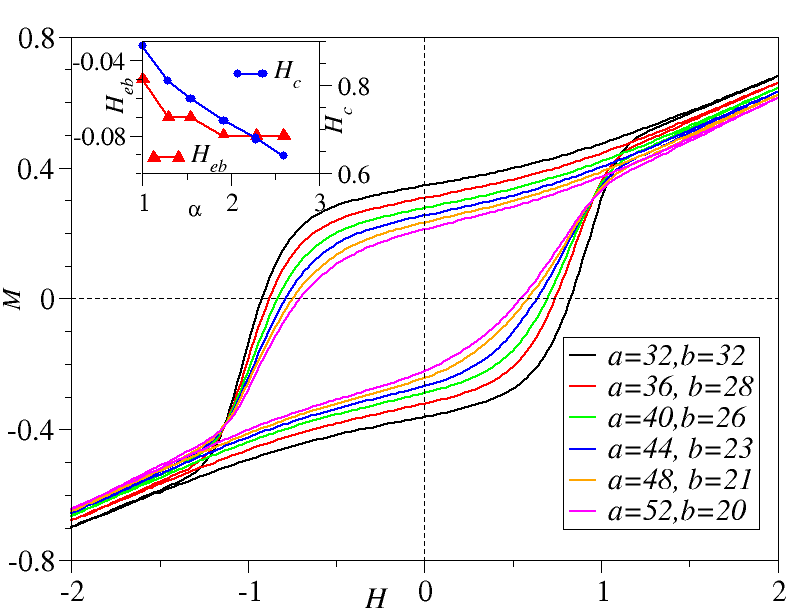}
	\caption{ (Color online) Hysteresis loop of elliptic core-shell structures, with approximately same area but  different  aspect 
	ratio $\alpha$=$a/b.$  All other parameters  are  taken from  Eq. (\ref{eq:default}). The inset shows  that the coercivity  $H_{c}$  decreases,  but $|H_{eb}|$  increases  with  	$\alpha.$ }
    \label{fig:aspect}
\par\end{centering}
\end{figure}

{\it Aspect ratio:}
      We  notice that  the  coercivity  of the elliptical  core-shell structure is   smaller  than that of the  circular one  
      with same area.  This indicates that  rod-like  structure  may have the smallest  coercivity, which  indeed has been observed 
      earlier \cite{Aditi}. Here we aim at studying  systematically,    how  aspect ratio   affects  $H_{eb}$ and  $H_c.$ To this end, 
      we  change the aspect ratio  $\alpha= \frac ab$   of the  ellipse  and follow   the  change in  its magnetic properties. Figure  \ref{fig:aspect} shows  the hysteresis loops;  the  coercivity  and exchange bias  are  plotted in the inset as a function of the aspect ratio.  The  interaction parameters  and the pinning parameters are taken  same as earlier. 
      Note,  that the coercive field  $H_c$    decreases, but    $|H_{eb}|$   increases  as  the aspect ratio $\alpha$  increases.

\section {Heterostructures}
The effects   of   surface pinning  are  also  expected to   be felt  in  heterostructures  at nano-scale. Heterostructures are layered magnetic composites; we  model  them as   $N_{F}$  number of   ferromagnetic layers placed on the top of  $N_{AF}$  number of   antiferromagnetic layers.   We have seen that    in  core-shell structure, where  core to shell  ratio    is  usually  high, the 
surface-pinning affects the system strongly when   the core   is antiferromagnetic and shell is ferromagnetic;  in other words  
affect of pinning  is stronger when  anti-ferro to ferro ratio is large.  Should  we expect the same here, i.e. whether the  surface pinning  would  affect the magnetic  properties of  heterostructures strongly when  $N_{AF}>N_{F} ?$
  \begin{figure}[h!]
 \begin{centering}
 	\includegraphics[scale=0.4]{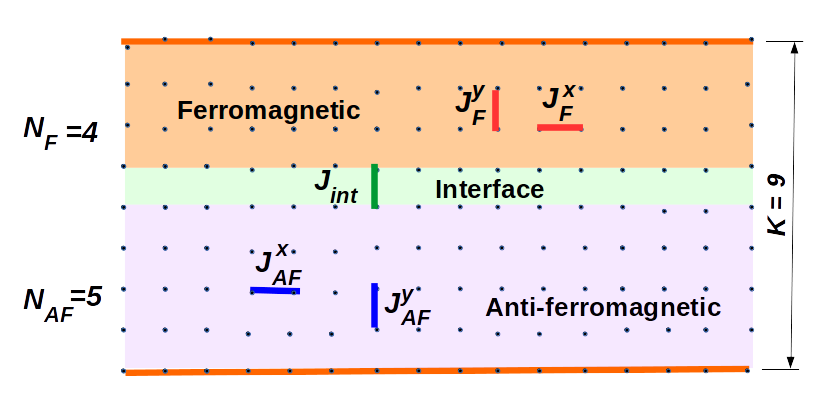}
 	\caption{ (Color online) Schematic representation of the heterostructure  of $K$ layers (each containing $L$ sites) of which $N_{AF}=5$ layer  are antiferromagnetic and  $N_{F}=4$  ferromagnetic. The intra-layer  coupling strengths   in ferro  and antiferro layers  are respectively $J^x_{F,AF}$ and  interlayer coupling strengths  are  $J^x_{F,AF}$.   The interaction at the  interface is $J_{int}.$
        }
 	\label{fig:heteroPic}
 	\par\end{centering}
\end{figure}

In two dimension,  the layers  can be modeled  by a line segment  of  length $L,$   lattice sites 
$i=1,2,\dots, L.$  The heterostructure  is a composite of  $N_{F}$   antiferromagnetic layers 
placed  one on the top of  o  $N_{AF}$  antiferromagnetic layers. 
The spins  of the  heterostructure are  denoted by  $s^k_i = \pm1,$  with $i=1,2,\dots, L,$  and  a layer index $k =1,2, \dots, K=N_{AF} + N_F$ (see Fig. \ref{fig:heteroPic}). The intralayer   interaction  strength of  Ising spins in ferromagnetic (antiferromagnetic) layers 
 are $J^x_F$  ($J^x_{AF}$),  whereas the    same in  interlayer are  $J^y_F$  ($J^y_{AF}$). 
At the interface  the interaction  strength   is $J_{int}$  which may  be positive (ferro)  or negative (antiferro).  
Corresponding Hamiltonian is, 
\bea
 {\cal H} &=&    -J^x_{AF} \sum_{k=1}^{N_{AF}}  \sum_{i=1}^{L-1} s^k_i s^k_{i+1}  
  -J^x_{F} \sum_{k=N_{AF}+1 }^{N_{F}}  \sum_{i=1}^{L-1} s^k_i s^k_{i+1} \cr
&& -J^y_{AF} \sum_{k=1}^{N_{AF}-1}  \sum_{i=1}^{L-1}s^k_i s^{k+1}_{i} 
  -J^y_F \sum_{k=N_{AF}+1}^{N_F-1}  \sum_{i=1}^{L-1} s^k_i s^{k+1}_{i} \cr
 &&  -J_{int} \sum_{i=1}^{L-1}s^{N_{AF}}_i s^{N_{AF}  +1}_{i}  -  H  \sum_{k=1}^{K}  \sum_{i=1}^{L}  s^k_i
 \label{eq:H_hetero}
 \eea
We also  consider  pinning of spins, which  occurs  at the top layer $k=K$  and the bottom layer $k=1.$   In  the  simulations,   starting  from a 
random  initial condition and fixed temperature $\beta^{-1} =1,$  we first increase the field from  $H=0$ to $H=H_{max}=2$  with a rate $0.02$ units per MCS. The zero filed hysteresis cycle is constructed now  by varying the field from  $H=2$  to $H=-2,$ and then back to $H=2.$ 
 
\begin{figure}[h!]
	\begin{centering}
		\includegraphics[scale=0.35]{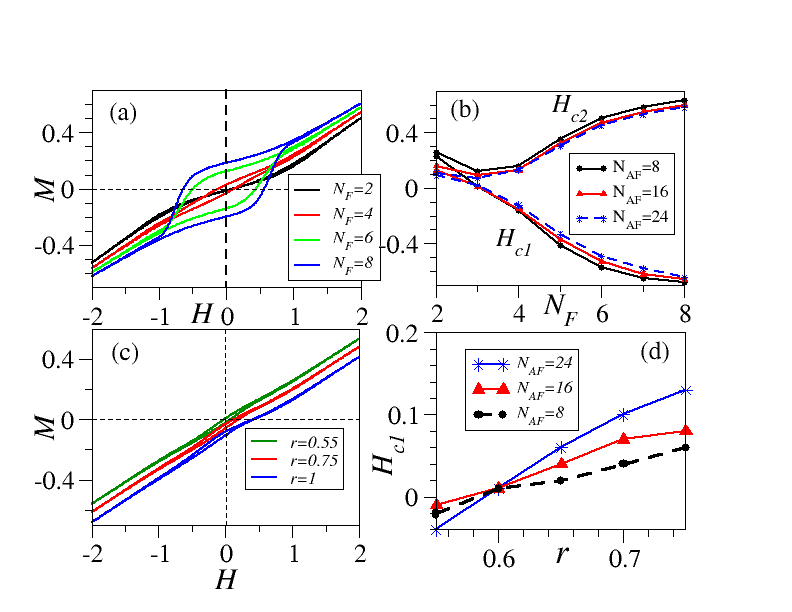}
		\par\end{centering}
	\caption{(Color online)  (a) Hysteresis loop of two-dimensional heterostructures  of $N_F$ ferromagnetic layers on the top of $N_{AF}=24$ antiferromagnetic layers. Here the length of each layer is $L=128,$  $\uparrow$-spin fraction $r=0.6.$   (b)  The dashed line shows   $H_{c1}$ and $H_{c2}$ as a function of $N_F.$ Similar curves for $N_{AF}= 16,8$ are also shown here (solid lines).
		(c) Hysteresis curves for different $r,$ for a  heterostructure with  $N_{AF}=8$  and $N_F=3.$  Corresponding  values of $H_{c1},$  as a function of $r,$  is  shown (dashed-line) in  (d). In   panel (d) we have also plotted  similar curves for   different  $N_{AF}=16,24.$  This indicates that for  a heterostructure with three  ferromagnetic layers ($N_F=3$), $H_{c1}$
		becomes positive when $\sim60$\%  of the pinned  spins are  $\uparrow.$ The  unspecified parameters here  are same as in Eq. (\ref{eq:default}).}	
	\label{fig:hetero}
\end{figure}

 First we calculate  the variation of  coercivity $H_c$ and exchange bias $H_{eb}$ by varying the  number of  ferromagnetic layers $N_F$  for a fixed number  of  antiferromagnetic layers $N_{AF}=24,$ as shown in  Fig. \ref{fig:hetero} (a).
 The  interaction parameters    are  taken to be   $J^x_{AF} = J^y_{AF}= J_{AF} =-0.5$,   $J^x_{F} = J^y_{F}= J_{F} =1.$  
At the  interface, we have  $J_{int} = 1.$  The  pinning parameters are, the  pinning density  $\eta =0.4$  and the   $\uparrow$-spin fraction   $r = 0.7$.  Corresponding    $H_{c1}$ and $H_{c2}$  are shown in   Fig. \ref{fig:hetero} (b)  in dashed line;  the solid lines there  correspond to $N_{AF}=8,16.$ In these  systems  we find  that the  coercivity  of  the heterostructure decreases monotonically as  $N_F$  decreases.
 
We also find an   interesting  finite size effect which  is  worth of mention.  In the  usual hysteresis cycle  the coercive  field  
$H_{c1}$  is negative. However,  in  heterostructure  studied here,  it appears that the    $H_{c1}$  can become  positive  
when    there are  three  or less   ferromagnetic layers  irrespective  of   antiferro-layers  $N_{AF}$ present.  It is clear from 
 Fig. \ref{fig:hetero} (b) that, for all three  values of  $N_{AF}=8, 16,24,$     the coercive  field  $H_{c1}$ becomes positive  when $N_F\le 3.$
 
 Does  this finite size effect originate from surface pinning ? To answer this, we  further study  the hysteresis  loop
 for  different  $r,$ the $\uparrow$-spin fraction.  Figure   \ref{fig:hetero} (c) shows  hysteresis curves for a heterostructure 
  of  $N_{AF}= 8$  and $N_F=3$  layers,  for different  $r=0.55, 0.75, 1.$. In all cases,   $H_{c1}$   is positive ($H_{c2}$ is 
  also positive  naturally).    For this  heterostructure,  variation of  $H_{c1}$  as a function of $r$ is shown  (dashed line) in 
 Fig.   \ref{fig:hetero} (d), which clearly  shows that    $H_{c1}$ crosses over  from  a negative  value to a positive one 
 at some  threshold  $r = r^* \simeq 0.6.$   In the same figure, we also plot   $H_{c1}$ versus $r$  for different  heterostructures  with three ferromagnetic layers, and different number of antiferro-layers $N_{AF}= 16,24;$   here  too $r^*\simeq0.6.$  In  fact,  the value  of $r^*$ depends  strongly on  the number of  ferromagnetic layers $N_{F}$ and its dependence of  antiferromagnetic  layers $N_{AF}$ is negligibly small. We find, 
 
\be
r^*  = \left\{  
\begin{matrix} 
 0 & N_F=2\cr
 \simeq 0.6 & N_F=3\cr
1 & N_F >3
 \end{matrix}
\right.
\ee

For $N_F=2$ $H_{c1}, H_{c2}$     are positive  for any  $r>0,$  whereas for $N_F > 3,$   coercive field  $H_{c1}$ is negative for any  $r>0.$
 
\section {Why  pinning affects hysteresis ? \label{sec:why}}
What we observe so far  from  the  Monte Carlo simulation of the core-shell nanostructure is that,  irrespective  of shape, size, value of the interaction strengh and pinning parameters, the exchange  bias $H_{eb}$ increases  with increase of pinning density $\eta$ and 
decrease with    $\uparrow$  spins fraction $r.$  To  understand this,  
we introduce a simple model of  the  surface ignoring   the interaction of the  surface spins  with  those  in the bulk (shell).
In two dimensional core-shell structure  of Ising spins,  the surface is  as a one dimensional chain  with periodic  boundary condition; the corresponding   Hamiltonian  is now, 

\bea
 {\cal H}_{surf}&=&  -J_{sh} \sum_{i=1}^L s_i s_{i+1},
 \label{eq: Hs}
 \eea
 where  $L$  is  total number of spins  on the  surface. For a circular core-shell structure studied here,   
 $L \simeq 2 \pi R$; in fact  a better approximation   is $L\simeq 4R,$ since   for every  $i \in  (-R,R)$  there are   two  boundary spins.
 We  assume that  $N$ spins on the surface  are pinned, of which $N_+$  spins are $\uparrow$, thus  $\sum_k S_k = 2N_+-N=M_b.$ Accordingly, 
 \be 
 \eta = \frac NL,  ~ r= \frac{N_+}N  ~ {\rm   and} ~   m_b\equiv \frac{M_b}{N} = 2 r -1. \label{eq:mb_r}
 \ee
 
 \begin{figure}[h!]
	\begin{centering}
		\hspace*{-2 cm} \includegraphics[width=12cm,height=3.5cm]{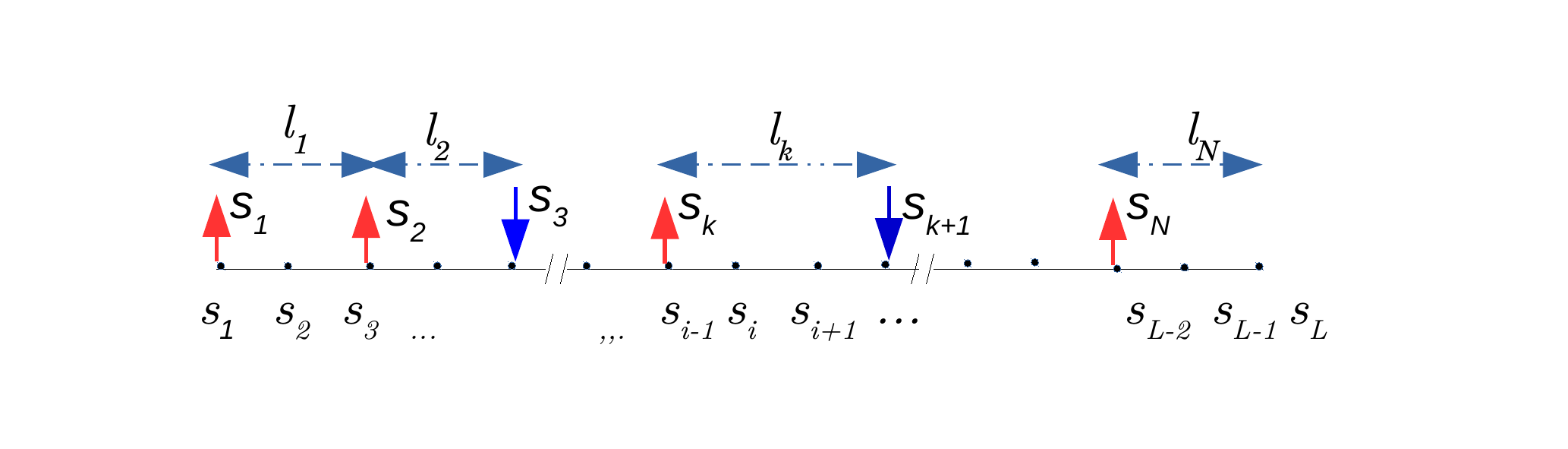}
				\par\end{centering} 
				\vspace*{-.5 cm}
	\caption{(Color online)  The pinned spins  are denoted  $\{S_1, S_2,\dots S_N\}.$ Here $S_1=s_1, S_2= s3, \dots S_N= s_{L-2}.$ Distance 
	between two consecutive  pinned spins $S_k$ and $S_{k+1}$  is $l_k.$ The periodic boundary condition ensures that  $s_{i+L}=s_i$ and $S_{k+N}=S_K.$}\label{fig:pinned_pic}
\end{figure}

 For notational  convenience, let us denote  the  pinned 
 spins as   $\{S_1, S_2,\dots S_N\}$ with  $S_1 = s_{i_{min}},$   where  $i_{min}$   is the  position index of the first pinned spin (see Fig. \ref{fig:pinned_pic}). As shown in the figure,  the separation   of  two  consecutive  pinned spins $S_k$ and  $S_{k+1}$  is  $l_k.$

The partition function   of the system can be written as 
\bea
Z_{L,N} (\{l_i\}) &=& \sum_{\{s_i\}}  e^{-\beta {\cal H}_{surf}} \cr &=& \prod_{k=1}^N   \langle S_k| T^{l_k}  |S_k\rangle   \delta \left(\sum_{k=1}^N S_k -M_b \right), \label{eq:partition}
\eea
where  $T=\begin{pmatrix} e^{K} &  e^{-K} \\  e^{-K}&   e^{K}\end{pmatrix}$ with   $K= \beta J_{sh}$ is the usual transfer matrix  of 
one-dimensional Ising model, and   $K= \beta J_{sh}.$

In presence of the constraint, that exactly $N_+$ out of $N$ pinned spins are $\uparrow$,  which is ensured by a   $\delta$-function  in above equation,  evaluating this partition-sum (\ref{eq:partition}) is difficult.  We  proceed to find a generating function of  $Z_{L,N} (\{l_i\}),$
\bea
{\cal Z}_L(x)  = \sum_{M_b=0}^\infty Z_{L,N} (\{l_i\}) x^{M_b} 
= Tr\left[ \prod_{k=1}^N  \left( T^{l_k} A \right)\right], \label{eq:ZL}\\
{\rm where}~~  A=  \sum_{S_k=\pm} x^{S_k} |S_k\rangle \langle S_k| = \begin{pmatrix}  x&0\\ 0&x^{-1} \end{pmatrix} 
\eea
Eigenvectors  of $T$  are  $|\pm\rangle=\begin{pmatrix} 1 \\ \pm 1  
\end{pmatrix}$
 with eigenvalues $\lambda_\pm=  e^{K} \pm  e^{-K},$ 
the generating function ${\cal Z}_L(x)$ can be evaluated   using the diagonalizing matrix  $U= \frac1 {\sqrt 2} \begin{pmatrix} 1&1\\1&-1 \end{pmatrix}.$
\bea
 {\cal Z}_L(x) 
=\lambda_+^L  (x+ \frac1x)^N  \left[ 1+  \left( \frac{ \lambda_-}{ \lambda_+}\right)^{l^*}  \left( \frac{x^2- 1}{x^2+1}\right)^2 +\dots \right] \label{eq:corr2}
\eea
where, we have used $\sum_{k=1}^N l_k =L$  and   $l^* = Min(\{l_i\})$ is the  smallest separation  between consecutive pinned spins. 

Note that, for any given choice of separations  $\{ l_k\},$  one can  calculate   ${\cal Z}_L(x)$ explicitly  using  Eq. (\ref{eq:corr2}).
In  the second step  here,  we  use a perturbation series  in  $\lambda= \frac{ \lambda_-}{ \lambda_+}= tanh(\beta J_{sh}),$ valid quite well 
in  large temperature  limit. The  dominant (zeroth order)  term  of   ${\cal Z}_L(x)$ does not  depend on    individual  separations $\{ l_k\},$ and the  next  order  correction  depends only on the smallest  separation $l^*.$ We have assumed that   the  smallest separation $l^*$ appear  only once in   $\{ l_k\};$   if  it  appears $n$ times (and  the separations  are  not  adjacent to each other)  then we have an additional  multiplicative factor $n,$ 
 \bea
{\cal Z}_L(x) & \simeq&  \lambda_+^L  (x+ x^{-1})^N  \left[ 1+ n \lambda ^{l^*}  \left( \frac{x^2- 1}{x^2+1}\right)^2 \right] 
\eea

The average  value of $M_b$  is  now 
\bea
\langle M_b\rangle &= &x  \frac{d}{dx} \ln  {\cal Z}_L(x)\cr &=& N  \frac{x^2- 1}{x^2+1} + \frac{ 4 n \lambda ^{l^*} x^2}{(1+x^2)^2 + n \lambda ^{l^*} (x^4-1)}
\label{eq:xM}
\eea
  
Further,   by redefining  $x= e^{-\beta h},$   one can check  from   Eq. (\ref{eq:ZL})   that ${\cal Z}_L(x)$ is the  partition function of  an effective  Hamiltonian 
\bea
 \tilde{\cal H}_{surf}&=&  -J_{sh} \sum_{i=1}^L s_i s_{i+1}  -h \sum_{k=1}^N S_k,
 \label{eq: Hs'}
 \eea
 where   the magnetic field $h$  is acting {\it selectively} only on the  pinned spins. 
 Thus the thermodynamic description of a system  having  $M_b$ number of excess $\uparrow$-spins among $N$ pinned ones,  is equivalent to 
 a system   without pinning, but an additional magnetic field $h$   acting  only on the  pinned spins. 
 The value of $h,$  for any given  $r,$ can be calculated  from  Eqs. (\ref{eq:xM})  and (\ref{eq:mb_r}) as 
 \be
r= \frac{1}{2} \left( 1+ \tanh(\beta h) + \frac{1}{N}  \frac{ n \lambda ^{l^*} {\rm sech}(\beta h)^2}{ 1- n \lambda ^{l^*} \tanh(\beta h)} \right). \label{eq:Mb_h}
 \ee
 This additional  field  on the  pinned spins,  along with  an  external magnetic field $H,$ produce an effective field  per site,
 \be
 H_{eff} = H+ \eta h  \label{eq:Heff}
 \ee
 In this system, the hysteresis   loop  would  be symmetric if the curves are plotted against $H_{eff}.$  But, if it is  plotted  against  the external field $H,$  the curves  would  become  asymmetric, in fact shifted  by  $\eta h.$
 Then, the exchange bias is  expected to be 
 \be
 H_{eb} =\eta h  \label{eq:Heb_h}
 \ee
 
In the  large $N$ limit,  it is  reasonable assume that the smallest  separation  between two consecutive spins is $l^*=1;$  in this limit  
Eq. (\ref{eq:Mb_h})  to a leading order in $N$ we have, 
\be
h= \frac 1\beta  \tanh^{-1}(m_b)  + \frac 1{\beta N} \frac{n \lambda }{n\lambda  m_b -1}. \label{eq:hh}
\ee
The effective field  $h,$ along with  Eq. (\ref{eq:Heb_h} ), indicates that  the exchange bias   increases linearly,  with pinning density $\eta$ and 
temperature $\beta^{-1};$ it also grows  monotonically  with  $m_b$ or $r= \frac{1+m_b}{2}.$  
Linear dependence  of  $H_{eb}$  on  $\eta, r$  are observed in   Fig.  \ref{fig:eta_and_r} (a) and (b) respectively. 
We have also  observed  linear   temperature dependence of $H_{eb}$  from  Monte Carlo simulations of  inverse core-shell 
nanoparticles (figures are not presented here).  

Note  that the interaction   strength $J_{sh},$  appears  in Eq. (\ref{eq:hh}) in the second term through  the relation 
$\lambda = tanh(\beta J_{sh}),$  is  suppressed  by  the factor $N.$  Thus, the   dependence of $H_{eb}$  on the 
interaction strength $J_{sh}$ is negligible. This is indeed  observed  from  simulations   results  described in 
Fig.  \ref{fig:eta_and_r} (c).

It is interesting that a simple  model  of the surface that   clearly ignore  the interaction of the   surface spins with other  
spins in the shell   reproduce  the properties of hysteresis qualitatively.

\section{Core-shell structures in three  dimension \label{sec:3D}}
In three spatial dimensions,  magnetic phase transition can  occur  in  systems  having discrete    (Ising) or continuous spins (like   XY and Heisenberg). In this section we extend our study to   core-shell structures   with spin interactions  given by either  Ising, XY  or   Heisenberg  models.  We  study  the  hysteresis   properties  of three  models, separately,   on  a  cubical core-shell with  core  ${\cal C}$  of length $R_c$ and  shell {$\cal S$} of width $R-R_c$, i.e., we have a cubical core  of size  $R_c^3$  and  particle size $R^3.$

\begin{figure}[h!]
	\begin{centering}
		\includegraphics[scale=0.3]{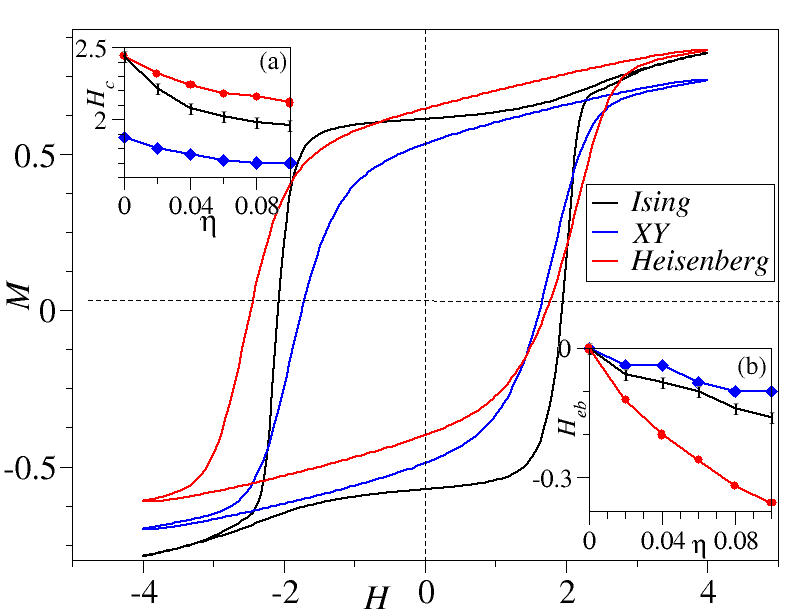}
		\par\end{centering}
	\caption{ (Color online) Hysteresis  loops  of a  cubical  core-shell structure (particle size $R^3$ and  core  size  $R_c^3$) for 
	Ising, XY and Heisenberg models, with   $R_c=26,$ $R=32,$  pinning density $\eta=0.1$ and   interaction parameters  $J_c = -0.5$, $J_{sh} = 1$, $J_{int} = 1.$  The insets (a) and (b) respectively shows  variation of  $H_c$ and $H_{eb}$ as a function of $\eta.$ 
	Heisenberg models generate maximum   exchange bias  $|H_{eb}|$ and  coercivity $H_c.$}
	\label{fig:3D}
\end{figure}

The  Hamiltonian  of the system is   now,
\bea
 {\cal H} &=&   -J_c \sum_{i\in {\cal C}, j\in {\cal C}} {\bf S}_i. {\bf S}_j  -J_{sh} \sum_{i\in {\cal S}, j \in {\cal S}}  {\bf S}_i. {\bf S}_j \cr &&-  J_{int}\sum_{i \in {\cal S}  j \in {\cal C}} {\bf S}_i. {\bf S}_j   - H \sum_{i\in {\cal C}, i\in {\cal S} } S^x_i,
 \label{eq:H3D}
 \eea
Now  spins ${\bf S}_i$ at  lattice site $i$ is a unit   vector, i.e.,  ${\bf S}_i. {\bf S}_i=1,$  and    `.'  corresponds the  
usual dot-product of vectors.  For Heisenberg model  ${\bf S}_i =( S^x_i,  S^y_i, S^z_i)$ has three components,  for  XY-model 
  ${\bf S}_i =( S^x_i,  S^y_i)$  has two components. For  the  Ising spins  ${\bf S}_i \equiv  S^x_i= \pm 1$   and    dot-product in Eq. (\ref{eq:H3D}) is  interpreted as simple multiplication.  The   direction of the magnetic field is taken as the  $x$-axis.

  We also consider  that   some of the spins  on the  surface  are pinned, and their  density is  $\eta$. In other words,  $\eta$ fraction of  total number of spins  on the   surface, having $6R^2  - 12 R +8$  sites in total, are pinned. We also assume that all the  pinned spins are aligned along   a  particular   direction, chosen to  be  $x$-direction.  Note  that   for core-shell  structure in two dimension,   surface pinning is characterized by two parameters, the pinning density   $\eta$ and  $\uparrow$-spin fraction $r.$  That   all  the pinned  spins  here   are parallel  to each other   is equivalent to  $r\equiv1,$ where    pinning effects are maximum.

   Now  we compare  the  hysteresis    curves  obtained from Monte Carlo simulations of  all the three models on a  core-shell structure with    $R_c=26$ and $R=32,$  which means that the width of the shell is $6$ lattice units.    The interaction parameters are taken as 
    $J_c = -0.5$, $J_{sh} = 1$, $J_{int} = 1.$  In Fig. \ref{fig:3D} we  have plotted the hysteresis curves  for  $\eta =0.4.$ 
   The insets (a)  and (b) in  Fig. \ref{fig:3D}  shows  respectively the variation of $H_c$ and $H_{eb}$   as a function of  
   pinning density $\eta.$     In  all  three  model, the coercivity $H_c$ goes  down  with increase of  $\eta,$ whereas as expected,  
   $|H_{eb}|$ increases.  For Heisenberg model, both  the  coercivity and     the change in   exchange  bias  is  quite high  compared  to   that of other models.

   Asymmetric   hysteresis, and thus  large   exchange bias, has been observed   earlier in core-shell  nanostructures.  For ferromagnetic core and antiferromagnetic   shell,  one  of  the primarily   factor that controls the magnetic properties  is the  core-shell 
   interaction   parameter.  Numerical studies of these  core-shell   structures  with Heisenberg  spin  interactions, additional 
   anisotropic  spin interaction and  crystal field  claims  that, under  field cooled conditions, the exchange bias  strongly depends  on  
    the nature and strength of  core-shell interaction \cite{Iglesias,Nogues}.  Here   we show that  surface pinning can affect  the exchange  bias strongly, even in the zero field cooled conditions and  in absence of  anisotropy  or crystal fields.   This pinning effect is  quite dominant 
    in inverse  core-shell  nanostructures  where  the core is antiferromagnetic and the shell, ferromagnetic.

\section{Discussions and Conclusion}

Wide  variation  in magnetic properties  has been observed in  nano-materials. In this mesoscopic scale,  the size, shape  and intrinsic 
spin   interactions  play a  significant role. This leads to  many fascinating finite size effects \cite{Nanotoday}  like,  magnetic anisotropy \cite{Jamet}, unusual magnetization \cite{Apsel} and  superparamagnetic behavior \cite{Arora}. In addition, external perturbations on nanoparticles, like   temperature, pressure  and  different solvent conditions  can tune the magnetic  properties in interesting  ways; organic  solvents  are known    agents which  may  pin  the spins on the surface \cite{Berkomitz}.

In this article we  study  how  magnetic properties are modulated  in  core-shell nanostructes and heterostructures  when   some of the  spins  on the surface get pinned. We find  that,  for core-shell nanoparticles, the hysteresis   behavior  is  significantly modified  when  the shell  is ferromagnetic and the core is antiferromagnetic, i.e., in a  inverse core-shell structure.  Ferromagnetic cores or antiferromagnetic shells  are  not affected  much by surface pinning.  Thus, we  primarily  focus on the inverse core-shell structure and study   hysteresis  in  zero filed cooled conditions  using  Monte  Carlo simulations.  We find that  the  exchange bias $H_{eb}$ changes linearly with   pinning density $\eta$  and it's  sign changes from being positive to negative    when  the   $\uparrow$-pinning fraction   $r$  is increased   beyond  $r=\frac12.$  The dependence  of  $H_{eb}$  on $J_{int}$ is found to be negligible. 

To  understand  these variations  we also propose  a  simple model of the surface  in \ref{sec:why}. For the  two dimensional   inverse core shell structure, the  surface is  an one dimensional ring, where Ising spins interact   with coupling strength $J_{sh}.$ We show that the  pinning  of  spins  can  generate an effective magnetic field in  the  positive (negative) direction when  the  fraction of $\uparrow$-spins $r$ 
is greater (smaller)  than $\frac12;$ accordingly, a  negative (positive)   exchange bias  $H_{eb}$  is  generated  in  the   in the  inverse core-shell structure.  We  show explicitly, in Eq. (\ref{eq:Heb_h}), that the  magnitude  $H_{eb}$ is  proportional to $\eta.$ 
The effective field   is  found  to be  independent of   number of pinned  spins $N$  in leading order and  it  varies as $\frac1 N$  as a next  order correction.  Interaction strength $J_{sh}$  appears only in the correction term which  is negligible for large $N.$
This simple model of surface, though  ignores the interaction of  spins on the surface with other spins in the shell,  explains the qualitative  properties of  the  hysteresis  quite  well. 

The following comment is in order. In  absence of  pinning,  core-shell structures  in three dimension,  under field cooled conditions, exhibit strong dependence  of exchange bias on  interface interaction $J_{int}$ \cite{Iglesias,Nogues} due to  uncompensated spins at the interface. Most models that study these behavior considered  the presence of spin anisotropic interactions and  crystal field  in the system.  In absence of  these complexities, in  the  present study  of  inverse core-shell structure we find a large exchange bias,  even in zero field cooled  condition  when surface spins are pinned. The    dependence of $H_{eb}$ on  $J_{int},$ however,   turns out to be  very weak.

To understand  how   
shell thickness affect the exchange bias and coercivity, we increase  the   core size $R_c$ for a  given $R.$  This  resulted  in  lower coercivity, but $|H_{eb}|$   is not affected much   since   the surface area  (where spin-pinning occurs)   does not change. 
Increase of $R$ and  $R_c$  proportionately, with  a fixed  $R_c/R,$  generates a thermodynamically large  core-shell structure.  In this case too the coercivity increases because  the  ferromagnetic shell is  larger in size, but the  exchange bias    decreases as because  the surface to volume ratio  decreases. Clearly,  the  asymmetry in the hysteresis  disappears in the thermodynamic limit indicating that the emergence of   exchange bias is only a mesoscopic phenomena.   

In addition, we also study  the  core-structures of  different shapes, namely triangular, square, circular, elliptical and some irregular shapes. Both,   the value  of  coercive field $H_c$  and exchange bias $|H_{eb}|$   changes   non-trivially. 
We separately investigate the role of aspect ratio  of an elliptical core-shell structure  and find  that the
coercivity $H_c$ decreases, but $|H_{eb}|$ increases with aspect ratio, re-confirming the fact  a  rod-like   nanostructure  may have the smallest  coercivity \cite{Aditi}.

We also investigate   the  nanoscale heterostructurs of  ferrromagnetic layers  grown on the  top of  antiferromagnetic  layers,  by  introducing   spin pinning in  the top and bottom surfaces. Like in core-shell structures, here too  the  exchange  bias  is  proportional to the  pinning density and   it changes sign  when  $\uparrow$-spin  fraction 
$r$   is increased beyond $\frac12.$  The coercivity, as expected,  depends  strongly on the number of ferromagnetic  layers.  An  interesting finite size effect is observed:   it turns out that there are  three or less ferromagnetic layers,  
the coercive   field  $H_{c1}$  changes  from being negative to  positive  when  $\uparrow$-spin  fraction crosses a threshold value 
$r^*;$  $r^*$ does not depend  on  the  number of antiferromagnetic layers.

In conclusion,  we study  in detail the   effect of  surface pinning  on  magnetic properties  of two dimensional  core-shell nanostructures  and  heterostructures in  nanoscale, where  spin interactions  are  considered Ising-like.  The pinning effect  turns out  to be more prominent  in a inverse core-shell structure where  spin interact antiferromagnetically  in the core  and ferromagnetically in the shell.
Studying   variation of hysteresis by changing  interaction and pinning parameters in 
Monte Carlo simulations we conclude    
that the exchange bias  increases  with   increase of  pinning density and the fraction of $\uparrow$-spins  which are  pinned.  
This behavior  could be explained from the analytical studies  of a model introduced here  that  captures only the  interaction of 
pinned  spins  on the surface. The shape  and size of the nanoparticles, which can be tuned experimentally,  also strongly   modify the  exchange bias and coercivity. The results  obtained  in two dimension, with Ising spins,  are found to be robust  in three dimension irrespective of whether the spins  interact following Ising, XY- or Heisenberg models.   We  believe that the new  mechanism of exchange bias  generated from the pinning of spins  on  the surface, whose  morphology and pinning density can be changed easily,   will help in  providing fruitful  technological applications in future.

\begin{acknowledgments}
AS acknowledges support from DST-INSPIRE in the form a Senior Research Fellowship. PKM acknowledges support from the Science and Engineering Research Board (SERB), India, Grant No. TAR/2018/000023.
\end{acknowledgments}


\begin{thebibliography}{99}
\bibitem{note1} To obtain hysteresis, the temperature must be  lower than the critical value; here  $\beta^{-1} =1$ is much smaller  than  the critical temperature  $T_c= \frac{2}{\ln(1+ \sqrt2)}$  of  Ising model on a  square lattice  with interaction strength  $J=1.$
\end{thebibliography}
\end{document}